\documentclass[a4paper,runningheads]{llncs}
\usepackage[latin1]{inputenc}
\usepackage[T1]{fontenc}
\usepackage{ae,aecompl}
\usepackage{amsmath,amsfonts,subfigure}
\usepackage[dvips]{graphicx}

\usepackage[dvips,colorlinks=false,bookmarks=false]{hyperref}

\usepackage[small,normal,bf,up]{caption2}

\hypersetup{
pdfauthor = {Lars Eirik Danielsen and Matthew G. Parker},
pdftitle = {Spectral Orbits and Peak-to-Average Power Ratio
of Boolean Functions with respect to the \{I,H,N\}\textasciicircum n Transform},
pdfkeywords = {Peak-to-Average Power Ratio, Quantum Error-Correcting Codes, 
Boolean Functions, Additive Codes over GF(4), Graph Theory, Local Complementation, Quantum Entanglement}
}

\DeclareMathOperator{\GF}{GF}

\newcommand{\cthl}[1]{\multicolumn{1}{c|}{#1}}
\begin{document}

\pagestyle{headings}

\mainmatter

\title{Spectral Orbits and Peak-to-Average Power Ratio
of Boolean Functions with respect to the $\{I,H,N\}^n$ Transform}
\titlerunning{Spectral Orbits and PAR of Boolean Functions w.r.t. $\{I,H,N\}^n$}
\author{Lars Eirik Danielsen \and Matthew G. Parker}
\institute{The Selmer Center, Department of Informatics, University of Bergen,\\ 
PB 7800, N-5020 Bergen, Norway\\
\texttt{\{\href{mailto:larsed@ii.uib.no}{larsed},\href{mailto:matthew@ii.uib.no}{matthew}\}@ii.uib.no}\\
\texttt{http://www.ii.uib.no/\~{}\{\href{http://www.ii.uib.no/~larsed}{larsed},\href{http://www.ii.uib.no/~matthew}{matthew}\}}
}

\maketitle

\begin{abstract}
We enumerate the inequivalent
self-dual additive codes over $\GF(4)$ of blocklength $n$,
thereby extending the sequence A090899 in \emph{The On-Line
Encyclopedia of Integer Sequences} from $n = 9$ to $n = 12$.
These codes have a well-known interpretation as quantum codes.
They can also be represented by graphs, where
a simple graph operation generates the orbits of equivalent codes.
We highlight the regularity and structure of some graphs that
correspond to codes with high distance.
The codes can also be interpreted as quadratic Boolean functions,
where inequivalence takes on a spectral meaning.
In this context we define PAR$_{IHN}$, 
peak-to-average power ratio with respect to the $\{I,H,N\}^n$ transform set.
We prove that PAR$_{IHN}$ of a Boolean function is
equivalent to the the size of the maximum independent set over
the associated orbit of graphs.
Finally we propose a construction technique to generate Boolean functions with
low PAR$_{IHN}$ and algebraic degree higher than 2. 
\end{abstract}

\section{Self-Dual Additive Codes over $\GF(4)$}

A quantum error-correcting code with parameters $[[n,k,d]]$ 
encodes $k$ qubits in a highly entangled state of $n$ qubits such that
any error affecting less than $d$ qubits can be detected, and any
error affecting at most $\frac{d-1}{2}$ qubits can be corrected.
A quantum code of the stabilizer type corresponds to a code 
$\mathcal{C} \subset \GF(4)^n$~\cite{Cald:Qua}. 
We denote $\GF(4) = \{0,1,\omega,\omega^2\}$, where $\omega^2 = \omega + 1$.
\emph{Conjugation} in $\GF(4)$ is defined by $\overline{x} = x^2$.
The \emph{trace map}, $\text{tr} : \GF(4) \mapsto \GF(2)$, is defined by
$\text{tr}(x) = x + \overline{x}$.
The \emph{trace inner product} of two vectors of length $n$ over $\GF(4)$, 
$\boldsymbol{u}$ and $\boldsymbol{v}$, is given by
$\boldsymbol{u} * \boldsymbol{v} = \sum_{i=1}^n tr(u_i \overline{v_i})$.
Because of the structure of stabilizer
codes, the corresponding code over $\GF(4)$, $\mathcal{C}$, will
be \emph{additive} and satisfy $\boldsymbol{u} * \boldsymbol{v} = 0$ for
any two codewords $\boldsymbol{u}, \boldsymbol{v} \in \mathcal{C}$.
This is equivalent to saying that the code must be \emph{self-orthogonal}
with respect to the trace inner product, i.e., $\mathcal{C} \subseteq \mathcal{C}^\perp$,
where $\mathcal{C}^\perp = \{ \boldsymbol{u} \in \GF(4)^n \mid
\boldsymbol{u}*\boldsymbol{c}=0, \forall \boldsymbol{c} \in \mathcal{C} \}$.

We will only consider codes of the special case where the dimension $k=0$.
Zero-dimensional quantum codes can be understood as highly-entangled
single quantum states which are robust to error. These codes map
to additive codes over $\GF(4)$ which are \emph{self-dual}~\cite{rains}, $\mathcal{C} = \mathcal{C}^\perp$.
The number of inequivalent
self-dual additive codes over $\GF(4)$ of blocklength $n$ has
been classified
by Calderbank~et~al.~\cite{Cald:Qua} for $n \le 5$,
by H\"{o}hn~\cite{Hohn:Klein} for $n \le 7$,
by Hein~et~al.~\cite{Hein:GrEnt} for $n \le 7$,
and by Glynn~et~al.~\cite{Glynn:Tome} for $n \le 9$.
Moreover, Glynn has recently posted these results as sequence A090899 in 
\emph{The On-Line Encyclopedia of Integer Sequences}~\cite{Slo:Seq}.
We extend this sequence from
$n = 9$ to $n = 12$ both for indecomposable and decomposable
codes as shown in table~\ref{tab:vncorbits}.
Table~\ref{tab:distances} shows the number of inequivalent indecomposable codes
by distance.
The distance, $d$, of a self-dual additive code over $\GF(4)$, $\mathcal{C}$,
is the smallest weight (i.e., number of nonzero components) of
any nonzero codeword in $\mathcal{C}$.
A database of orbit representatives with information about
orbit size, distance, and weight distribution is also available~\cite{database}.

\begin{table}
\centering
\caption{Number of Inequivalent Indecomposable ($i_n$) and (Possibly) Decomposable
($t_n$) Self-Dual Additive Codes Over $\GF(4)$}
\label{tab:vncorbits}
\begin{tabular}{|c||c|c|c|c|c|c|c|c|c|c|c|c|}
\hline
$n$ & 1 & 2 & 3 & 4 & 5 & 6 & 7 & 8 & 9 & 10 & 11 & 12 \\
\hline
$i_n$ & 1 & 1 & 1 & 2 &  4 & 11 & 26 & 101 & 440 & 3,132 & 40,457 & 1,274,068 \\
$t_n$ & 1 & 2 & 3 & 6 & 11 & 26 & 59 & 182 & 675 & 3,990 & 45,144 & 1,323,363 \\
\hline
\end{tabular}
\end{table}

\begin{table}
\centering
\caption{Number of Indecomposable Self-Dual Additive Codes Over $\GF(4)$ by Distance}
\label{tab:distances}
\begin{tabular}{|c||r|r|r|r|r|r|r|r|r|r|r|}
\hline
$d \backslash n$ & \cthl{2} & \cthl{3} & \cthl{4} & \cthl{5} & \cthl{6} & 
\cthl{7} & \cthl{8} & \cthl{9} & \cthl{10} & \cthl{11} & \cthl{12} \\
\hline
2     & 1 & 1 & 2 & 3 &  9 & 22 &  85 & 363 & 2,436 & 26,750 &   611,036 \\
3     &   &   &   & 1 &  1 &  4 &  11 &  69 &   576 & 11,200 &   467,513 \\
4     &   &   &   &   &  1 &    &   5 &   8 &   120 &  2,506 &   195,455 \\
5     &   &   &   &   &    &    &     &     &       &      1 &        63 \\
6     &   &   &   &   &    &    &     &     &       &        &         1 \\
\hline
Total & 1 & 1 & 2 & 4 & 11 & 26 & 101 & 440 & 3,132 & 40,457 & 1,274,068 \\
\hline
\end{tabular}
\end{table}

\section{Graphs, Boolean Functions, and LC-Equivalence}

A self-dual additive code over $\GF(4)$
corresponds to a \emph{graph state}~\cite{Hein:GrEnt} if its generator matrix, $G$, 
can be written as $G = \Gamma + \omega I$, where $\Gamma$ is a symmetric matrix over $\GF(2)$ with
zeros on the diagonal. The matrix $\Gamma$ can be interpreted as the adjacency
matrix of a simple undirected graph on $n$ vertices. It has been shown
by Schlingemann and Werner~\cite{Sch:QG}, Grassl~et~al.~\cite{Gras:QG},
Glynn~\cite{Glynn:Graph}, and Van~den~Nest~et~al.~\cite{VanD:Gr}
that all stabilizer states can be
transformed into an equivalent graph state. Thus all
self-dual additive codes over $\GF(4)$ can be represented by graphs.
These codes also have another interpretation
as quadratic Boolean functions over $n$ variables. A quadratic
function, $f$, can be represented by an adjacency matrix, $\Gamma$,
where $\Gamma_{i,j} = \Gamma_{j,i} = 1$ if $x_ix_j$ occurs in $f$,
and $\Gamma_{i,j} = 0$ otherwise.

\begin{figure}
 \centering
 \subfigure[The ``Wheel'']
 {\hspace{25pt}\includegraphics[width=.26\linewidth]{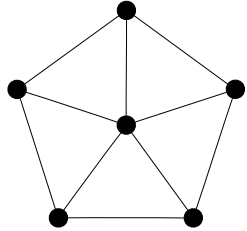}\hspace{25pt}\label{wheel}}
 \subfigure[The ``2-clique of 3-cliques'']
 {\hspace{25pt}\includegraphics[width=.26\linewidth]{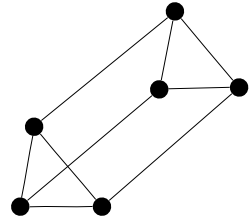}\hspace{25pt}\label{doubleclique}}
 \caption{The LC Orbit of the [[6,0,4]] ``Hexacode''}
 \label{hexacode}
\end{figure}

\begin{example}
A self-dual additive code over $\GF(4)$ with parameters $[[6,0,4]]$ is
generated by the generator matrix
\[
\begin{pmatrix}
  \omega & 0 & 0 & 1 & 1 & 1\\
  0 & \omega & 0 & \omega^2 & 1 & \omega\\
  0 & 0 & \omega & \omega^2 & \omega & 1\\
  0 & 1 & 0 & \omega & \omega^2 & 1\\
  0 & 0 & 1 & \omega & 1 & \omega^2\\
  1 & \omega^2 & 0 & \omega & 0 & 0
\end{pmatrix}.
\]
We can transform the generator matrix into 
the following generator matrix of an equivalent code 
corresponding to a graph state,
\[
\begin{pmatrix} 
  \omega & 0 & 0 & 1 & 1 & 1 \\ 
  0 & \omega & 1 & 1 & 0 & 1 \\ 
  0 & 1 & \omega & 1 & 1 & 0 \\ 
  1 & 1 & 1 & \omega & 1 & 1 \\ 
  1 & 0 & 1 & 1 & \omega & 0 \\ 
  1 & 1 & 0 & 1 & 0 & \omega 
\end{pmatrix}
= \Gamma + \omega I.
\]
$\Gamma$ is the adjacency matrix of the graph shown in fig.~\ref{wheel}.
It can also be represented by the quadratic Boolean function 
$f(\boldsymbol{x}) = x_0x_3+x_0x_4+x_0x_5+x_1x_2+x_1x_3+x_1x_5+x_2x_3+x_2x_4+x_3x_4+x_3x_5$.
\end{example}

Recently, Glynn~et~al.~\cite{Glynn:Tome,Glynn:Graph} has re-formulated the
primitive operations that map equivalent self-dual additive codes over
$\GF(4)$ to each other as a single, primitive operation on the associated
graphs. This symmetry operation is referred to as \emph{Vertex Neighbourhood Complementation}
(VNC). It was also
discovered independently by Hein~et~al.~\cite{Hein:GrEnt}
and by Van den Nest~et~al.~\cite{VanD:Gr}. The
identification of this problem as a question of establishing the
\emph{local unitary equivalence} between those quantum states that can
be represented as graphs or Boolean functions was presented by
Parker and Rijmen at SETA'01~\cite{Par:QE}. Graphical representations have
also been identified in the
context of quantum codes by Schlingemann and Werner~\cite{Sch:QG} and
by Grassl~et~al.~\cite{Gras:QG}. 
VNC is another name for \emph{Local Complementation} (LC), referred
to in the context of \emph{isotropic systems} by Bouchet~\cite{Bouch:Iso,Bouch:VNC}.
LC is defined as follows.
\begin{definition}
Given a graph $G=(V,E)$ and a vertex $v \in V$. Let $N_v \subset V$ 
be the neighbourhood of $v$, i.e., the set of vertices adjacent to $v$. 
The subgraph induced by $N_v$ is complemented to obtain the LC image $G^v$.
\end{definition}
It is easy to verify that $(G^v)^v = G$. 
\begin{theorem}[Glynn et~al.~\cite{Glynn:Tome,Glynn:Graph}]
Two graphs $G$ and $H$ correspond
to equivalent self-dual additive codes over $\GF(4)$ iff there is a finite sequence of vertices
$v_1, v_2, \ldots, v_s$, such that $(((G^{v_1})^{v_2})^{\cdots})^{v_s} = H$.
\end{theorem}

The symmetry rule can also be described in terms of quadratic Boolean functions.
\begin{definition}
If the quadratic monomial $x_ix_j$ occurs in the algebraic normal form of the
quadratic Boolean function $f$,
then $x_i$ and $x_j$ are mutual neighbours in the graph represented by $f$,
as described by the $n \times n$ symmetric adjacency matrix,
$\Gamma$, where $\Gamma_{i,j} = \Gamma_{j,i} = 1$ if $x_ix_j$ occurs in $f$,
and $\Gamma_{i,j} = 0$ otherwise.
The quadratic Boolean functions $f$ and $f'$ are \emph{LC equivalent} if
\[
f'(\boldsymbol{x}) = f(\boldsymbol{x}) + \sum_{\substack{j,k \in N_a \\ j < k}} x_jx_k \pmod{2},
\]
where $a \in \mathbb{Z}_n$ and $N_a$ comprises the neighbours of $x_a$ in the graph
representation of $f$.
\end{definition}
A finite number of repeated applications of the LC operation generates 
the orbit classes presented in this paper and, therefore, induces an equivalence between
quadratic Boolean functions. We henceforth refer to this equivalence
as \emph{LC-equivalence} and the associated orbits as \emph{LC orbits}.
If the graph representations of two self-dual additive codes over
$\GF(4)$ are isomorphic, they are also considered to be equivalent.
This corresponds to a permutation of the labels of the vertices in the
graph or the variables in the Boolean function. We only count members
of an LC orbit up to isomorphism. 
As an example, fig.~\ref{hexacode}
shows the graph representation of the two only non-isomorphic members
in the orbit of the $[[6,0,4]]$ ``Hexacode''.

A recursive algorithm, incorporating the package \emph{nauty}~\cite{nauty} 
to check for graph isomorphism, was
used to generate the LC orbits enumerated in table~\ref{tab:vncorbits}.
Only the LC orbits of indecomposable codes (corresponding to
connected graphs) were generated, since all decomposable codes 
(corresponding to unconnected graphs) can easily be constructed by
combining indecomposable codes of shorter lengths.

Consider,
(a) self-dual additive codes over $\GF(4)$ of blocklength $n$,
(b) pure quantum states of $n$ qubits which are joint eigenvectors of a
commuting set of operators from the Pauli Group~\cite{Cald:Qua},
(c) quadratic Boolean functions of $n$ variables, (d) undirected graphs 
on $n$ vertices. Then, under a suitable interpretation, we consider objects 
(a), (b), (c), and (d) to be mathematically identical.

\section{Regular Graph Structures}

\begin{figure}
 \centering
 \subfigure[The {$[[6,0,4]]$} \newline ``2-clique of 3-cliques'']
 {\hspace{3pt}\includegraphics[width=.31\linewidth]{S04-Danielsen-Parker-2cl3cl}\hspace{3pt}\label{2cl3cl}}
 \subfigure[The {$[[12,0,6]]$} \newline ``3-clique of 4-cliques'']
 {\hspace{3pt}\includegraphics[width=.31\linewidth]{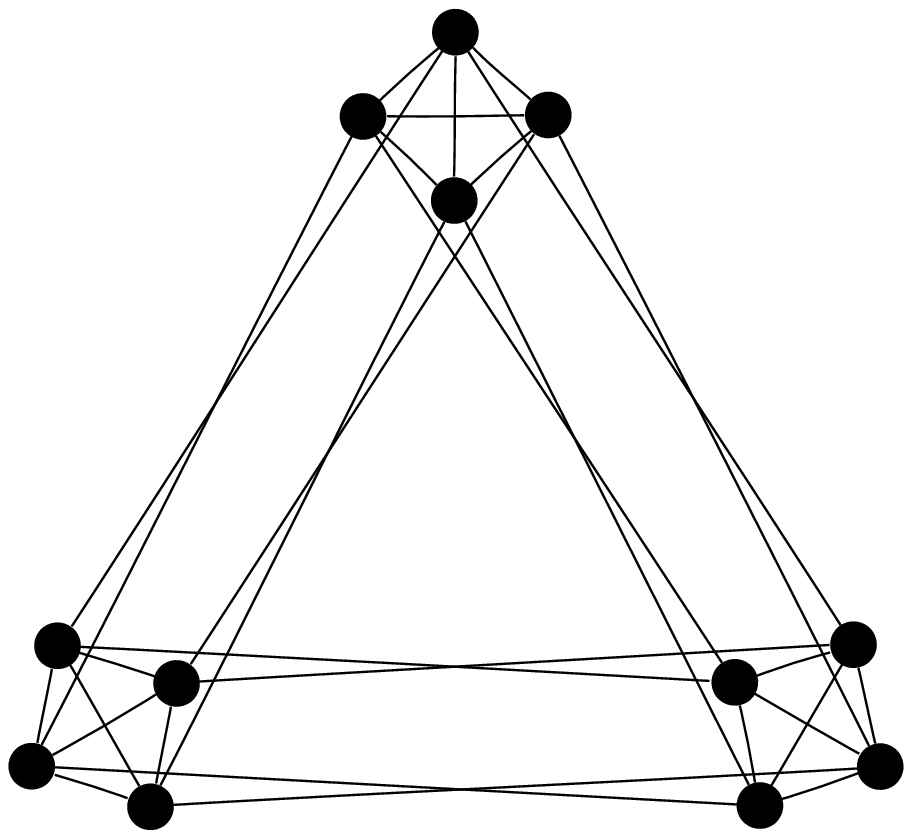}\hspace{3pt}\label{3cl4cl}}
 \subfigure[The {$[[18,0,6]]$} \newline ``2-clique of 3-cliques of 3-cliques'']
 {\hspace{3pt}\includegraphics[width=.31\linewidth]{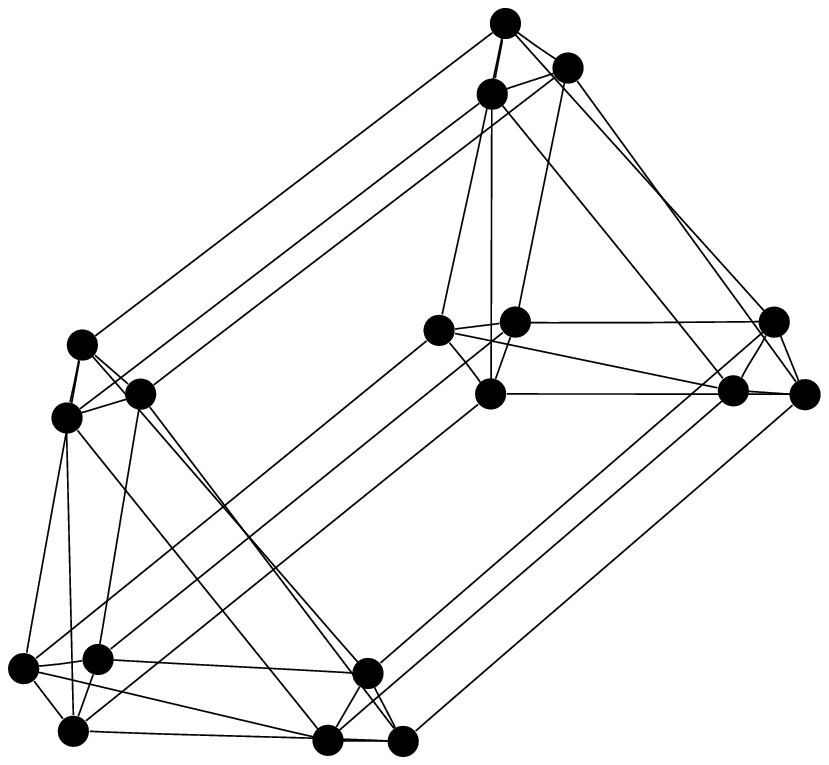}\hspace{3pt}\label{2cl3cl3cl}}\\
 \subfigure[The {$[[20,0,8]]$} \newline ``5-clique of 4-cliques'']
 {\hspace{3pt}\includegraphics[width=.41\linewidth]{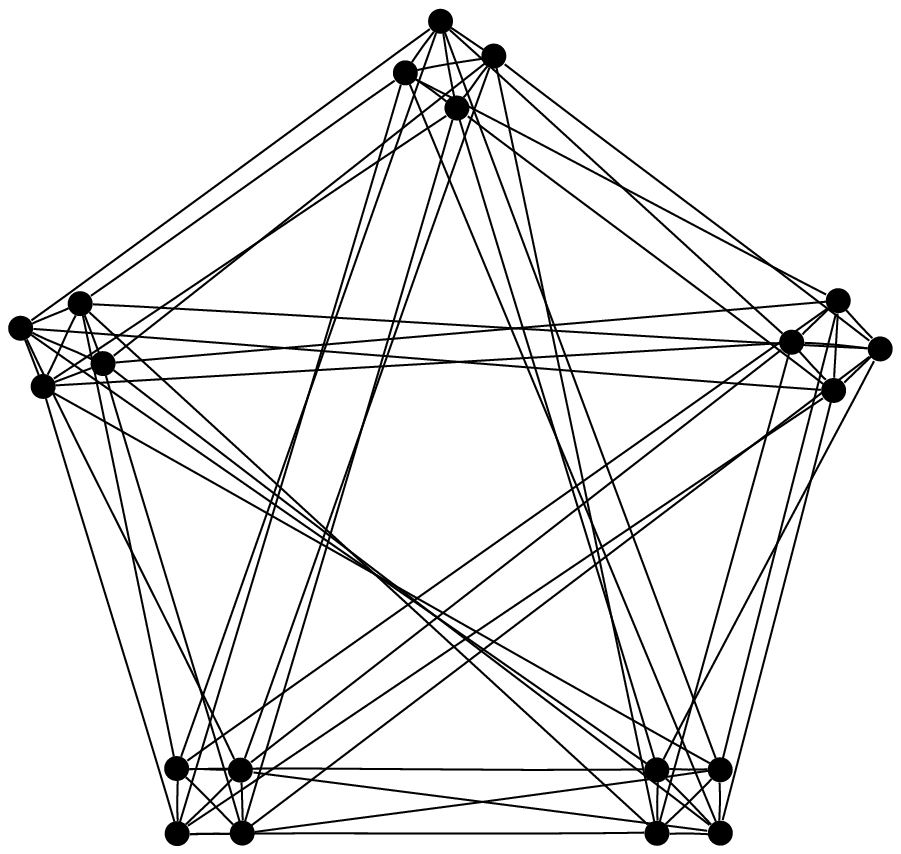}\hspace{3pt}\label{5cl4cl}}
 \subfigure[The {$[[25,0,8]]$} \newline ``5-clique of 5-cliques'']
 {\hspace{3pt}\includegraphics[width=.41\linewidth]{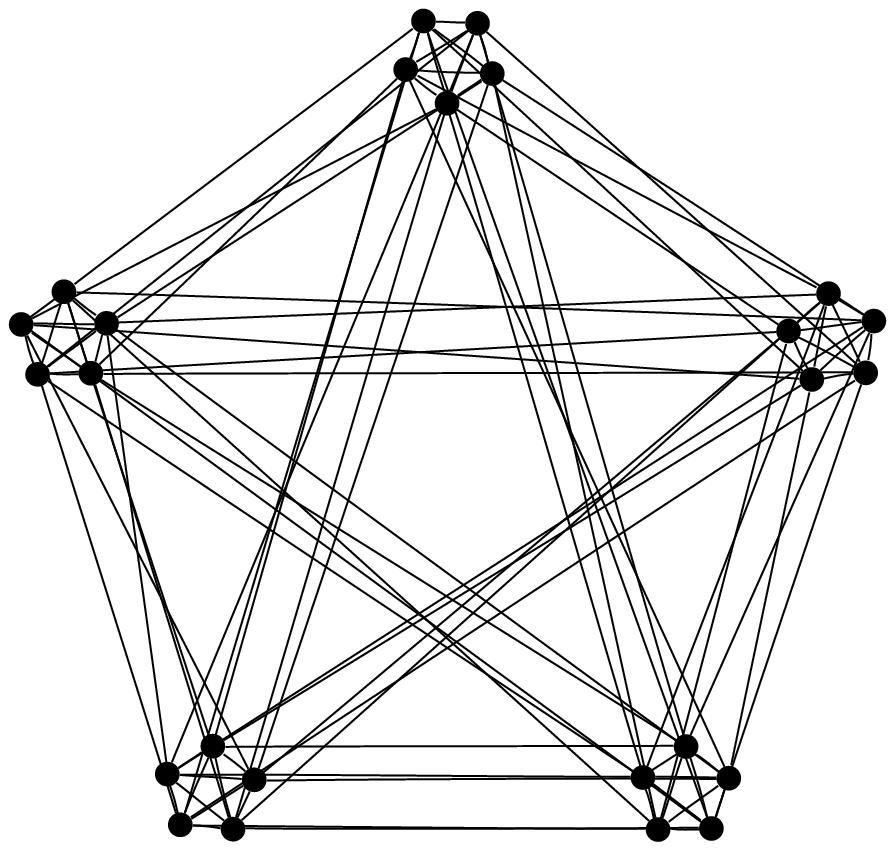}\hspace{3pt}\label{5cl5cl}}
 \caption{``Nested Clique'' Graphs}
 \label{graphs}
\end{figure}

Although a number of constructions for self-dual
additive codes over $\GF(4)$ exist~\cite{Glynn:Tome,Gull:Circ}, it appears that 
the underlying symmetry of
their associated graphs has not been identified or exploited
to any great extent. We highlight the
regularity and structure of some graphs that correspond to 
self-dual additive codes over $\GF(4)$ with high distance.
Of particular interest are the highly regular ``nested clique''
graphs.
Fig.~\ref{graphs} shows a few examples of such graphs.
There is an upper bound on the possible distance of self-dual additive codes
over $\GF(4)$~\cite{rains}. Codes that meet this bound are called \emph{extremal}.
Other bounds on the distance also exist~\cite{Cald:Qua,Gras:QECCs}.
Of the codes corresponding to graphs shown in fig.~\ref{graphs},
the $[[6,0,4]]$, $[[12,0,6]]$, and $[[20,0,8]]$ codes are extremal.
To find the ``nested clique'' graph representations, one may
search through the appropriate LC orbits. Also note that
all ``nested clique'' graphs we have identified so far have
\emph{circulant} adjacency matrices. An exhaustive search of
all graphs with circulant adjacency matrices of up to 30 vertices has been performed.

If $d$ is the distance of a self-dual additive code over $\GF(4)$,
then every vertex in the corresponding graph must have a
vertex degree of at least $d - 1$.
This follows from the fact that
a vertex with degree $\delta$ corresponds to a row in
the generator matrix, and therefore a codeword, of weight $\delta + 1$.
All the graphs shown in fig.~\ref{graphs} satisfy the minimum possible 
regular vertex degree for the given distance.
Some extremal self-dual additive codes over $\GF(4)$ do not have any regular graph representation, for example 
the unique $[[11,0,5]]$ and $[[18,0,8]]$ codes.
For codes of length above 25 and distance higher than 8 the graph structures 
get more complicated. For example, with a non-exhaustive search,
we did not find a graph representation of a 
$[[30,0,12]]$ code with regular vertex degree lower than 15.

\section{The $\{I,H,N\}^n$ Transform}

LC-equivalence between two graphs can be interpreted as an equivalence
between the generalised Fourier spectra of the two associated Boolean
functions. 
\begin{definition}
Let
\[
I =
\begin{pmatrix}
1 & 0 \\
0 & 1
\end{pmatrix},\quad
H = {\frac{1}{\sqrt{2}}}
\begin{pmatrix}
1 & 1 \\
1 & -1
\end{pmatrix},\quad
N = {\frac{1}{\sqrt{2}}}
\begin{pmatrix}
1 & i \\
1 & -i
\end{pmatrix},
\]
where $i^2 = -1$, be the Identity, Hadamard, and Negahadamard kernels, respectively.
\end{definition}
These are \emph{unitary} matrices, i.e.,
$II^{\dag} = HH^{\dag} = NN^{\dag} = I$, where $\dag$ means
\emph{conjugate transpose}.
Let $f$ be a Boolean function on
$n$ variables and $\boldsymbol{s} = 2^{-\frac{n}{2}}(-1)^{f(\boldsymbol{x})}$
be a vector of length $2^n$.
Let $s_j$, where $j \in \mathbb{Z}_{2^n}$, be the $j$th coordinate of $\boldsymbol{s}$.
Let $U = U_0 \otimes U_1 \otimes \cdots \otimes U_{n-1}$ where
$U_k \in \{I,H,N\}$, and $\otimes$ is the \emph{tensor product} (or 
\emph{Kronecker product}) defined as
\[
A \otimes B 
= \begin{pmatrix}a_{00}B & a_{01}B & \cdots \\ a_{10}B & a_{11}B & \cdots \\
\vdots & \vdots & \ddots \end{pmatrix}.
\]
Let $\boldsymbol{S} = U\boldsymbol{s}$ for any of the $3^n$
valid choices of the $2^n \times 2^n$ transform $U$.
Then the set of $3^n$ vectors, $\boldsymbol{S}$, is a multispectra
with respect to the transform set, $U$, with $3^n2^n$ spectral points.
We refer to this multispectra as
the spectrum with respect to the $\{I,H,N\}^n$ transform.
(Using a similar
terminology, the spectrum with respect to the $\{H\}^n$ transform would
simply be the well-known Walsh-Hadamard spectrum).
It can be shown
that the $\{I,H,N\}^n$ spectrum of an LC orbit is invariant to within
coefficient permutation. Moreover if, for a specific choice of $U$, $\boldsymbol{S}$ is
flat (i.e., $|S_i| = |S_j|$, $\forall i,j$),
then we can write $\boldsymbol{S} = v^{4f'(\boldsymbol{x}) + h(\boldsymbol{x})}$, where $f'$ is
a Boolean function, $h$ is any function from $\mathbb{Z}_2^n$ to $\mathbb{Z}_8$, and
$v^4 = -1$.
If the algebraic degree of $h(\boldsymbol{x})$ is $\le 1$, we
can always eliminate $h(\boldsymbol{x})$ by post-multiplication
by a tensor product of matrices from
$\mathcal{D}$, the set of $2 \times 2$ diagonal and
anti-diagonal unitary matrices~\cite{vncbent}, an operation
that will never change the spectral coefficient magnitudes.
Let $M$ be the multiset of $f'$
existing within the $\{I,H,N\}^n$ spectrum for the subcases where
$h(\boldsymbol{x})$ is of
algebraic degree $\le 1$. The \emph{$\{I,H,N\}^n$-orbit}
of $f$ is then the set of distinct members of $M$. In particular, if $f$
is quadratic then the $\{I,H,N\}^n$-orbit is the LC orbit~\cite{vncbent}.

\begin{example}
We look at the function $f(\boldsymbol{x}) = x_0x_1 + x_0x_2$.
The corresponding bipolar vector, ignoring the normalization factor, is
\[
\boldsymbol{s} = (-1)^{f(\boldsymbol{x})} = (1,1,1,-1,1,-1,1,1)^T.
\]
We choose the transform $U = N \otimes I \otimes I$ and get the
result
\[
\boldsymbol{S} = U\boldsymbol{s} = (v,v^7,v^7,v,v^7,v,v,v^7)^T, \quad v^4=-1.
\]
We observe that $|S_i| = 1$, $\forall i$, which means that $\boldsymbol{S}$
is flat and can be expressed as
\[
\boldsymbol{S} =  v^{4(x_0x_1+x_0x_2+x_1x_2)+(6x_0+6x_1+6x_2+1)}.
\]
We observe that $h(\boldsymbol{x})$, the terms that are not divisible by 4, 
are all linear or constant. We can therefore eliminate $h(\boldsymbol{x})$,
in this case by using the transform
\[
D = \begin{pmatrix}1 & 0 \\ 0 & i\end{pmatrix} \otimes 
    \begin{pmatrix}1 & 0 \\ 0 & i\end{pmatrix} \otimes
    \begin{pmatrix}v^7 & 0 \\ 0 & v\end{pmatrix}.
\]
We get the result
\[
D\boldsymbol{S} = (-1)^{x_0x_1 + x_0x_2 + x_1x_2},
\]
and thus $f'(\boldsymbol{x}) = x_0x_1 + x_0x_2 + x_1x_2$.
The functions $f$ and $f'$ are in the same $\{I,H,N\}^n$ orbit, and since
they are quadratic functions, the same LC orbit. This can be verified
by applying the LC operation to the vertex corresponding to the
variable $x_0$ in the graph representation of either function.
\end{example}

\section{Peak-to-Average Power Ratio w.r.t. $\{I,H,N\}^n$}

\begin{definition}
The peak-to-average power ratio of a vector, $\boldsymbol{s}$,
with respect to the
$\{I,H,N\}^n$ transform~\cite{Par:SB} is 
\[
\text{\emph{PAR}}_{IHN}(\boldsymbol{s}) = 
2^n \max_{\substack{ \forall U \in \{I,H,N\}^n \\ 
\forall k \in \mathbb{Z}_{2^n} }} |S_k|^2, \quad \text{where $\boldsymbol{S} = U\boldsymbol{s}$.}
\]
\end{definition}
If a vector, $\boldsymbol{s}$, has a completely flat $\{I,H,N\}^n$ spectrum
(which is impossible) then
PAR$_{IHN}(\boldsymbol{s}) = 1$. If $\boldsymbol{s} = 2^{-\frac{n}{2}}(1,1,\ldots,1,1)$ then
PAR$_{IHN}(\boldsymbol{s}) = 2^n$. A typical vector, $\boldsymbol{s}$, will have a PAR$_{IHN}(\boldsymbol{s})$
somewhere between these extremes.
For quadratic functions, PAR$_{IHN}$ will always be a power of 2.
The PAR of $\boldsymbol{s}$ can be alternatively
expressed in terms of the \emph{generalised nonlinearity}~\cite{Par:SB}, 
\[
\gamma(f) = 2^{\frac{n}{2} - 1} \left( 2^{\frac{n}{2}}
  - {\sqrt{\text{PAR}_{IHN} \left( \boldsymbol{s} \right) }} \right),
\]
but in this paper we use the PAR measure.
Let $\boldsymbol{s} = 2^{-\frac{n}{2}}(-1)^{f(\boldsymbol{x})}$, as before. When we talk about the
PAR$_{IHN}$ of $f$ or its associated graph $G$,
we mean PAR$_{IHN}(\boldsymbol{s})$. It is desirable to find
Boolean functions with high generalised nonlinearity and therefore
low PAR$_{IHN}$~\cite{Dan:APC}.
PAR$_{IHN}$ is an invariant of the $\{I,H,N\}^n$ orbit and, in particular,
the LC orbit.
We observe that Boolean functions from LC orbits associated
with self-dual additive codes over $\GF(4)$ with high
distance typically have low PAR$_{IHN}$. This is not surprising
as the distance of a quantum code has been shown
to be equal to the
recently defined \emph{Aperiodic Propagation Criteria distance} 
(APC distance)~\cite{Dan:APC} of the associated quadratic Boolean function,
and APC is derived from the aperiodic autocorrelation which is, in turn,
the autocorrelation ``dual'' of the spectra with respect to $\{I,H,N\}^n$.
Table~\ref{QuadPARs} shows PAR$_{IHN}$ values for every LC orbit
representative for $n \le 12$.

\begin{table}
\centering
\caption{PAR$_{IHN}$ of LC Orbit Representatives}
\label{QuadPARs}
\begin{tabular}{|c||r|r|r|r|r|r|r|r|r|r|r|r|r|}
\hline
$n$ & \multicolumn{11}{|c|}{Number of orbits with specified PAR$_{IHN}$} \\
& \cthl{2} & \cthl{4} & \cthl{8} & \cthl{16} & \cthl{32} & \cthl{64} & \cthl{128} 
& \cthl{256} & \cthl{512} & \cthl{1024} & \cthl{2048} \\
\hline
1   & 1 &   &   &        &         &         &        &       &     &    & \\
2   & 1 &   &   &        &         &         &        &       &     &    & \\
3   &   & 1 &   &        &         &         &        &       &     &    & \\
4   &   & 1 & 1 &        &         &         &        &       &     &    & \\
5   &   & 1 & 2 &      1 &         &         &        &       &     &    & \\
6   &   & 1 & 5 &      4 &       1 &         &        &       &     &    & \\
7   &   &   & 6 &     14 &       5 &       1 &        &       &     &    & \\
8   &   &   & 9 &     52 &      32 &       7 &      1 &       &     &    & \\
9   &   &   & 2 &    156 &     212 &      60 &      9 &     1 &     &    & \\
10  &   &   & 1 &    624 &   1,753 &     639 &    103 &    11 &   1 &    & \\
11  &   &   &   &  3,184 &  25,018 &  10,500 &  1,578 &   163 &  13 &  1 & \\ 
12  &   &   &   & 12,323 & 834,256 & 380,722 & 43,013 & 3,488 & 249 & 16 & 1 \\
\hline
\end{tabular}
\end{table}

\begin{definition}
Let $\alpha(G)$ be the independence number of a graph $G$, i.e., the size
of the maximum independent set in $G$. Let $[G]$ be the set of all graphs in
the LC orbit of $G$.
We then define $\lambda(G) = \max_{H \in [G]} \alpha(H)$, i.e., the
size of the maximum independent set over all graphs in the LC orbit of $G$.
\end{definition}
Consider as an example the Hexacode
which has two non-isomorphic graphs in its orbit (see fig.~\ref{hexacode}).
It is evident that the size of the largest independent set of each graph is 2, so $\lambda = 2$.
The values of $\lambda$ for all LC orbits for $n \le 12$ clearly show that
$\lambda$ and $d$, the distance of the associated self-dual additive code over $\GF(4)$, are related. 
LC orbits associated with codes with high distance typically have small values for $\lambda$.
Table~\ref{MaxIS} summarises this observation by giving the ranges of $\lambda$ observed for 
all LC orbits associated with codes of given lengths and distances. For instance, $[[12,0,2]]$ codes exist with
any value of $\lambda$ between 4 and 11, while $[[12,0,5]]$ and $[[12,0,6]]$ codes only exist
with $\lambda = 4$.

\begin{table}
\centering
\caption{Range of Maximum Independent Set Size}
\label{MaxIS}
\begin{tabular}{|c||c|c|c|c|c|c|c|c|c|c|c|c|c|}
\hline
$d$ & \multicolumn{11}{|c|}{Range of $\lambda$ for specified $n$} \\
                      & 2 & 3 & 4 &  5      & 6      & 7     & 8      & 9      & 10      & 11    & 12 \\ 
\hline
                    2 & $\boldsymbol{1}$ & $\boldsymbol{2}$ & $\boldsymbol{2}$,3 & 3,4    & 3--5   & 3--6  & 3--7   & 4--8   & 4--9    & 4--10 & 4--11\\
                    3 &   &   &     & $\boldsymbol{2}$      & 3      & $\boldsymbol{3}$,4   & 3,4    & 3--5   & 4--6    & 4--7  & 4--8  \\
                    4 &   &   &     &        & $\boldsymbol{2}$      &       & $\boldsymbol{3}$,4    & $\boldsymbol{3}$,4    & $\boldsymbol{3}$--5    & 4--6  & 4--7  \\
                    5 &   &   &     &        &        &       &        &        &         & $\boldsymbol{4}$     & 4  \\
                    6 &   &   &     &        &        &       &        &        &         &       & $\boldsymbol{4}$  \\
\hline
\end{tabular}
\end{table}

\begin{definition}
Let $\Lambda_n$ be the minimum value of $\lambda$ over all LC orbits with $n$ vertices.
\end{definition}
From table~\ref{MaxIS} we observe that
$\Lambda_n = 2$ for $n$ from 3 to 6, $\Lambda_n = 3$ for $n$ from 7 to 10, and
$\Lambda_n = 4$ when $n$ is 11 or 12. 
\begin{theorem}
$\Lambda_{n+1} \ge \Lambda_n$, i.e.,
$\Lambda_n$ is monotonically nondecreasing when the number of vertices is increasing.
\end{theorem}
\begin{proof}
Consider a graph $G=(V,E)$ with $n+1$ vertices.
Select a vertex $v$ and let $G'$ be the induced subgraph on
the $n$ vertices $V \backslash \{v\}$.
We generate the LC-orbit of $G'$. The LC operations may add or remove
edges between $G'$ and $v$, but the presence of $v$ does not affect the 
LC orbit of $G'$.
The size of the largest independent set in the LC orbit of $G'$ is at least
$\Lambda_n$. This is also an independent set in the LC orbit of $G$, so
$\Lambda_{n+1} \ge \Lambda_n$.\qed
\end{proof}

A very loose lower bound on $\Lambda_n$ can also be given.
Consider a graph containing a clique of size $k$. It is easy to see that
an LC operation on any vertex in the clique will produce an independent set of size $k-1$.
Thus the maximum clique in an LC orbit, where the largest independent set has size $\lambda$,
can not be larger than $\lambda + 1$.
If $r$ is the \emph{Ramsey number} $R(k,k+1)$~\cite{ramsey}, then it is guaranteed that all simple undirected
graphs with minimum $r$ vertices will have either an independent set of size $k$ or a clique of size $k+1$.
It follows that all LC orbits with at least $r$ vertices must have $\lambda \ge k$.
Thus $\Lambda_n \ge k$ for $n \ge r$.
For instance, $R(3,4)=9$, so LC orbits with at least 9 vertices can not have $\lambda$ smaller than 3.

For $n>12$, we have computed the value of $\lambda$ for some graphs corresponding to self-dual additive 
codes over $\GF(4)$ with high distance. This gives us upper bounds on the value of $\Lambda_n$, 
as shown in table~\ref{lambda}. The bounds on $\Lambda_{13}$ and $\Lambda_{14}$ are tight, 
since $\Lambda_{12} = 4$ and $\Lambda_{n+1} \ge \Lambda_n$.

\begin{table}
\centering
\caption{Upper Bounds on $\Lambda_n$}
\label{lambda}
\begin{tabular}{|c||c|c|c|c|c|c|c|c|c|c|c|c|}
\hline
$n$             & 13 & 14 & 15 & 16 & 17 & 18 & 19 & 20 & 21 \\
\hline
$\Lambda_n \le$ &  4 &  4 &  5 &  5 &  5 &  6 &  6 &  6 &  9 \\
\hline
\end{tabular}
\end{table}

\begin{figure}
 \centering
 \includegraphics[height=50pt]{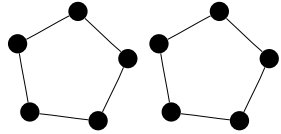}
 \includegraphics[height=50pt]{S04-Danielsen-Parker-5circle}
 \caption{The ``Double 5-Cycle'' Graph}
 \label{fivecircle}
\end{figure}

For $n=10$, there is a unique LC orbit that satisfies, optimally, $\lambda = 3$,
PAR$_{IHN} = 8$ and $d = 4$.
One of the graphs in this orbit is the \emph{graph complement} of the
``double 5-cycle'' graph, shown in fig.~\ref{fivecircle}.

\begin{theorem}[Parker and Rijmen~\cite{Par:QE}]\label{th:ihindset}
Given a graph $G=(V,E)$ with a maximum independent set $A \subset V$, 
$|A| = \alpha(G)$. Let $\boldsymbol{s} = (-1)^{f(\boldsymbol{x})}$,
where $f(\boldsymbol{x})$ is the boolean function representation of $G$.
Let $U = \bigotimes_{i \in A} H_i \bigotimes_{i \not\in A} I_i$, i.e., the
transform applying $H$ to variables corresponding to vertices $v \in A$
and $I$ to all other variables.
Then $\max_{\forall k \in \mathbb{Z}_{2^n}} |S_k|^2 = 2^{\alpha(G)}$, where 
$\boldsymbol{S} = U\boldsymbol{s}$.
\end{theorem}

Arratia~et~al.~\cite{interlace2} introduced the \emph{interlace polynomial} $q(G,z)$
of a graph~$G$. Aigner and van der Holst~\cite{aigner} later introduced the 
interlace polynomial $Q(G,z)$.
Riera and Parker~\cite{spectralinterlace} showed that $q(G,z)$ is related
to the $\{I,H\}^n$ spectra of the quadratic boolean function corresponding to $G$,
and that $Q(G,z)$ is related to the $\{I,H,N\}^n$ spectra.

\begin{theorem}[Riera and Parker~\cite{spectralinterlace}]\label{th:degree}
Let $f$ be a quadratic boolean function and $G$ its associated graph.
Then PAR$_{IHN}$ of $f$ is equal to $2^{\deg Q(G,z)}$, where
$\deg Q(G,z)$ is the degree of the interlace polynomial $Q(G,z)$.
\end{theorem}

\begin{theorem}\label{th:indset}
If the maximum independent set over all graphs in the LC orbit $[G]$ has size $\lambda(G)$,
then all functions corresponding to graphs in the orbit will have PAR$_{IHN} = 2^{\lambda(G)}$.
\end{theorem}
\begin{proof}
Let us for brevity define $P(G) = \text{PAR}_{IHN}(\boldsymbol{s})$,
where $\boldsymbol{s} = 2^{-\frac{n}{2}}(-1)^{f(\boldsymbol{x})}$, and
$f(\boldsymbol{x})$ is the boolean function representation of $G$.
From theorem~\ref{th:ihindset} it follows that $P(G) \ge 2^{\lambda(G)}$.
Choose $H=(V,E) \in [G]$ with $\alpha(H) = \lambda(G)$. If $|V| = 1$ or 2, the
theorem is true. We will prove the theorem for $n > 2$ by induction on $|V|$.
We will show that $P(H) \le 2^{\alpha(H)}$, which is equivalent to saying that
$P(G) \le 2^{\lambda(G)}$.
It follows from theorem~\ref{th:degree} and the definition of $Q(H,z)$ by 
Aigner and van der Holst~\cite{aigner} that
$P(H) = \max \{P(H\backslash u)$, $P(H^u \backslash u)$, $P(((H^u)^v)^u \backslash u)\}$.
(We recall that $H^u$ denotes the LC operation on vertex $u$ of $H$.)
Assume, by induction hypothesis, that $P(H \backslash u) = 2^{\lambda(H \backslash u)}$.
Therefore, $P(H\backslash u) = 2^{\alpha(K\backslash u)}$ for some
$K\backslash u \in [H\backslash u]$. Note that
$K\backslash u \in [H\backslash u]$ implies $K \in [H]$.
It must then be true that
$\alpha(K\backslash u) \le \alpha(K) \le \alpha(H)$,
and it follows that $P(H\backslash u) \le 2^{\alpha(H)}$.
Similar arguments hold for $P(H^u \backslash u)$ and $P(((H^u)^v)^u \backslash u)$,
so $P(H) \le 2^{\alpha(H)}$.\qed
\end{proof}
As an example, the Hexacode has $\lambda=2$ and therefore PAR$_{IHN} = 2^2 = 4$.

\begin{corollary}
Any quadratic Boolean function on $n$ or more variables must 
have PAR$_{IHN} \ge 2^{\Lambda_n}$.
\end{corollary}

\begin{definition}
PAR$_{IH}$ is the peak-to-average power ratio with respect to the
transform set $\{I,H\}^n$, otherwise defined in the same way as PAR$_{IHN}$.
\end{definition}
\begin{definition}
PAR$_{l}$ is the peak-to-average power ratio with respect to the infinite transform 
set $\{U\}^n$, consisting of matrices of the form
\[
U = \begin{pmatrix}
\cos \theta & \sin \theta e^{i\phi}\\
\sin \theta & -\cos \theta e^{i\phi}
\end{pmatrix},
\]
where $i^2 = -1$, and $\theta$ and $\phi$ can take any real values.
$\{U\}$ comprises all $2 \times 2$ unitary transforms to within
a post-multiplication by a matrix from $\mathcal{D}$, 
the set of $2 \times 2$ diagonal and anti-diagonal unitary matrices. 
\end{definition}

\begin{theorem}[Parker and Rijmen~\cite{Par:QE}]\label{th:bipartite}
If $\boldsymbol{s}$ corresponds to a bipartite graph, then
PAR$_{l}(\boldsymbol{s})$ = PAR$_{IH}(\boldsymbol{s})$.
\end{theorem}
It is obvious that $\{I,H\}^n \subset \{I,H,N\}^n \subset \{U\}^n$,
and therefore that PAR$_{IH} \le \text{PAR}_{IHN} \le \text{PAR}_{l}$.
We then get the following corollary of theorems~\ref{th:indset} and \ref{th:bipartite}.
\begin{corollary}
If an LC orbit, $[G]$, contains a bipartite graph, then
all functions corresponding to graphs in the orbit will have PAR$_{l} = 2^{\lambda(G)}$.
\end{corollary}
Thus, all LC orbits with a bipartite member have PAR$_{IHN} = $ PAR$_{l}$.
Note that these orbits will always have PAR$_{l} 
\ge 2^{\left\lceil\frac{n}{2}\right\rceil}$~\cite{Par:QE}
and that the fraction of LC orbits which have a bipartite member appears to decrease
exponentially as the number of vertices increases.
In the general case, PAR$_{IHN}$ is only a lower bound on PAR$_{l}$. For example,
the Hexacode has PAR$_{IHN} = 4$, but a tighter lower bound on PAR$_l$ is $4.486$~\cite{Par:QE}.
(This bound has later been improved to $5.103$~\cite{wheelPAR}.)

\section{Construction for Low PAR$_{IHN}$}

So far we have only considered \emph{quadratic} Boolean functions which correspond
to graphs and self-dual additive codes over $\GF(4)$.
For cryptographic purposes, we are interested in Boolean functions of degree higher
than $2$.
Such functions can be represented by \emph{hypergraphs}, but
they do not correspond to quantum stabilizer codes or self-dual additive codes over $\GF(4)$.
A non-quadratic Boolean function, $f(\boldsymbol{x})$, can, however, be interpreted as 
a quantum state described by the probability distribution vector 
$\boldsymbol{s} = 2^{-\frac{n}{2}}(-1)^{f(\boldsymbol{x})}$.
A single quantum state corresponds to a quantum code of dimension zero whose
distance is the APC distance~\cite{Dan:APC}. The APC distance is
the weight of the minimum weight quantum error operator that
gives an errored state not orthogonal to the original state 
and therefore not guaranteed to be detectable.

We are interested in finding Boolean functions of algebraic degree greater than 2 with
low PAR$_{IHN}$, but exhaustive searching becomes infeasible with more than a few variables.
We therefore propose a construction technique for 
nonquadratic Boolean functions with low PAR$_{IHN}$ using
the best quadratic functions as building blocks.
Before we describe our construction we must first state what we mean
by ``low PAR$_{IHN}$''. 
For $n = 6$ to $n = 10$ we computed PAR$_{IHN}$ for samples
from the space $\mathbb{Z}_2^{2^n}$,
to determine the range of PAR$_{IHN}$ we can expect just by guessing.
Table~\ref{PARSamples} summarises these results. 
If we can
construct Boolean functions with PAR$_{IHN}$ lower than the sampled minimum,
we can consider our construction to be somewhat successful.

\begin{table}
\centering
\caption{Sampled Range of PAR$_{IHN}$ for $n = 6$ to $10$}
\label{PARSamples}
\begin{tabular}{|c||c|c|c|c|c|}
\hline
 n                     & 6          & 7            & 8              & 9               & 10   \\
\# samples             & 50000      & 20000        & 5000           & 2000            & 1000 \\
\hline
Range of PAR$_{IHN}$  & 6.5--25.0 & 9.0--28.125 & 12.25--28.125 & 14.0625--30.25 & 18.0--34.03 \\
\hline 
\end{tabular}
\end{table}

\begin{figure}
 \centering
 \scalebox{0.88}{\includegraphics{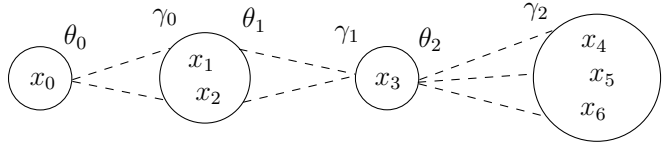}}
 \caption{Example of Construction with PAR$_{HN} \le 8$}
 \label{construction1}
\end{figure}

Parker and Tellambura~\cite{Par:PAR1,Par:LowPAR} 
proposed a generalisation of the Maiorana-McFarland construction for
Boolean functions that satisfies a tight upper bound on PAR with respect
to the $\{H,N\}^n$ transform (and other transform sets),
this being a form of Golay Complementary
Set construction and a generalisation of the construction of
Rudin and Shapiro and of Davis and Jedwab~\cite{Dav:PF}.
Let $p(\boldsymbol{x})$ be a Boolean function on
$n = \sum_{j=0}^{L-1} t_j$
variables, where $T = \{t_0,t_1,\ldots,t_{L-1}\}$ is a set of positive
integers and
$\boldsymbol{x} \in \mathbb{Z}_2^n$. Let $\boldsymbol{y_j} \in \mathbb{Z}_2^{t_j}$, $0 \le j < L$,
such that $\boldsymbol{x} = \boldsymbol{y_0} \times \boldsymbol{y_1} \times \cdots \times \boldsymbol{y_{L-1}}$.
Construct $p(\boldsymbol{x})$ as follows.
\begin{equation}
p(\boldsymbol{x}) = \sum_{j=0}^{L-2} \theta_j(\boldsymbol{y_j})\gamma_j(\boldsymbol{y_{j+1}})
 + \sum_{j=0}^{L-1} g_j(\boldsymbol{y_j}) \label{eq2},
\end{equation}
where $\theta_j$ is a
permutation: $\mathbb{Z}_2^{t_j} \rightarrow \mathbb{Z}_2^{t_{j+1}}$,
$\gamma_j$ is a
permutation: $\mathbb{Z}_2^{t_{j+1}} \rightarrow \mathbb{Z}_2^{t_j}$,
and $g_j$ is any Boolean function on $t_j$ variables.
It has been shown~\cite{Par:LowPAR} that the function $p(\boldsymbol{x})$ will
have PAR$_{HN} \le 2^{t_{\text{max}}}$, where
$t_{\text{max}}$ is the largest integer in $T$.
It is helpful to visualise this construction graphically, as in fig.~\ref{construction1}.
In this example, the size of the largest partition is $3$, so PAR$_{HN} \le 8$, regardless
of what choices we make for $\theta_j$, $\gamma_j$, and $g_j$.

Observe that if we set $L=2$, $t=t_0=t_1$, let $\theta_0$ be the identity permutation,
and $g_0 = 0$,
construction~(\ref{eq2}) reduces to the Maiorana-McFarland construction over $2t$ variables.
Construction~(\ref{eq2}) can also be viewed as a generalisation of the ``path graph'', 
$f(\boldsymbol{x}) = x_0x_1 + x_1x_2 + \cdots + x_{n-2}x_{n-1}$, 
which has optimal PAR with respect to $\{H,N\}^n$.
Unfortunately, the ``path graph'' is not a particularly good construction for low PAR$_{IHN}$. But
as we have seen, graphs corresponding to self-dual additive codes over $\GF(4)$ with
high distance do give us Boolean functions with low PAR$_{IHN}$.
We therefore propose the following generalised construction.
\begin{equation}
p(\boldsymbol{x}) =
\sum_{i=0}^{L-1}\sum_{j=i+1}^{L-1} \Gamma_{i,j}(\boldsymbol{y_i})\Gamma_{j,i}(\boldsymbol{y_j})
 + \sum_{j=0}^{L-1} g_j(\boldsymbol{y_j}) \label{eq3},
\end{equation}
where $\Gamma_{i,j}$ is either a
permutation: $\mathbb{Z}_2^{t_i} \rightarrow \mathbb{Z}_2^{t_j}$, or
$\Gamma_{i,j} = 0$,
and $g_j$ is any Boolean function on $t_j$ variables.
It is evident that $\Gamma$ can be thought of as a
``generalised adjacency matrix'', where the entries,
$\Gamma_{i,j}$, are no longer
0 or 1 but, instead, 0 or permutations from
$\mathbb{Z}_2^{t_i}$ to $\mathbb{Z}_2^{t_j}$. Construction~(\ref{eq2}) then becomes a
special case where $\Gamma_{i,j} = 0$ except for when $j = i + 1$ (i.e., the ``generalised
adjacency matrix'' of the ``path graph'').
In order to minimise PAR$_{IHN}$ we choose the form of the matrix $\Gamma$
according to the adjacency matrix of a self-dual additive code
over $\GF(4)$ with high distance.
We also choose the ``offset'' functions, $g_j$, to be
Boolean functions corresponding to self-dual additive codes
over $\GF(4)$ with high distance.
Finally for the non-zero $\Gamma_{i,j}$ entries,
we choose selected permutations,
preferably nonlinear to increase the overall degree.
Here are some initial results which demonstrate that, using~(\ref{eq3}),
we can construct
Boolean functions of algebraic degree greater than 2 with low PAR$_{IHN}$.
(We use an abbreviated ANF notation for some many-term Boolean functions, e.g.
$012,12,0$ is short for $x_0x_1x_2 + x_1x_2 + x_0$.)

\begin{figure}
 \centering
 \scalebox{0.88}{\includegraphics{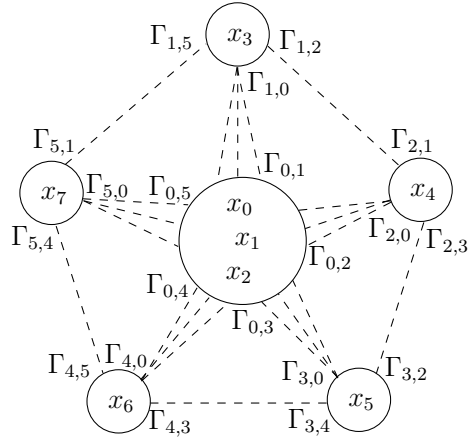}}
 \caption{Example of Construction with low PAR$_{IHN}$}
 \label{construction2}
\end{figure}

\begin{example}[$n=8$]
Use the Hexacode graph
$f = 01,$ $02,$ $03,$ $04,$ $05,$ $12,$ $23,$ $34,$ $45,$ $51$ as a template.
Let $t_0 = 3$, $t_1 = t_2 = t_3 = t_4 = t_5 = 1$. (See fig.~\ref{construction2}.)
We use the following matrix $\Gamma$.
\[
\Gamma = \begin{pmatrix}
0\hspace{3pt} & \hspace{3pt}02,1\hspace{3pt} & \hspace{3pt}02,1\hspace{3pt} & 
  \hspace{3pt}02,1\hspace{3pt} & \hspace{3pt}02,1\hspace{3pt} & \hspace{3pt}02,1\\
3 & 0 & 3 & 0 & 0 & 3\\
4 & 4 & 0 & 4 & 0 & 0\\
5 & 0 & 5 & 0 & 5 & 0\\
6 & 0 & 0 & 6 & 0 & 6\\
7 & 7 & 0 & 0 & 7 & 0
\end{pmatrix}
\]
Let $g_0(\boldsymbol{y_0}) = 01,$ $02,$ $12$
and all other $g_j$ any arbitrary affine functions.
Then, using~(\ref{eq3}) to construct $p(\boldsymbol{x})$ we get
$p(\boldsymbol{x}) = 023,$ $024,$ $025,$ $026,$ $027,$ $01,$ $02,$ $12,$ $13,$ 
 $14,$ $15,$ $16,$ $17,$ $34,$ $37,$ $45,$ $56,$ $67$.
Then $p(\boldsymbol{x})$ has PAR$_{IHN} = 9.0$.
\end{example}

\begin{example}[$n=8$]
Use the Hexacode graph
$f = 01,$ $02,$ $03,$ $04,$ $05,$ $12,$ $23,$ $34,$ $45,$ $51$ as a template.
Let $t_0 = 3$, $t_1 = t_2 = t_3 = t_4 = t_5 = 1$. (See fig.~\ref{construction2}.)
We use the following matrix $\Gamma$.
\[
\Gamma = \begin{pmatrix}
0\hspace{3pt} & \hspace{3pt}02,1\hspace{3pt} & \hspace{3pt}12,0,1,2\hspace{3pt} & 
  \hspace{3pt}01,02,12,1,2\hspace{3pt} & \hspace{3pt}01,02,12\hspace{3pt} & \hspace{3pt}02,12,1,2\\
3 & 0 & 3 & 0 & 0 & 3\\
4 & 4 & 0 & 4 & 0 & 0\\
5 & 0 & 5 & 0 & 5 & 0\\
6 & 0 & 0 & 6 & 0 & 6\\
7 & 7 & 0 & 0 & 7 & 0
\end{pmatrix}
\]
Let $g_0(\boldsymbol{y_0}) = 01,12$
and all other $g_j$ any arbitrary affine functions.
Then, using~(\ref{eq3}) to construct $p(\boldsymbol{x})$ we get
$p(\boldsymbol{x}) = 015,$ $016,$ $023,$ $025,$ $026,$ $027,$ $124,$ $125,$ $126,$ $127,$ 
$01,$ $04,$ $12,$ $13,$ $14,$ $15,$ $17,$ $24,$ $25,$ $27,$ $34,$ $37,$ $45,$ $56,$ $67$.
Then $p(\boldsymbol{x})$ has PAR$_{IHN} = 9.0$.
\end{example}

\begin{figure}
 \centering
 \scalebox{0.88}{\includegraphics{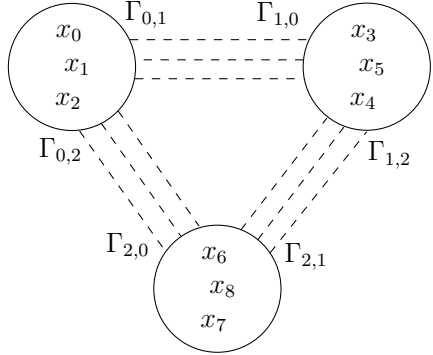}}
 \caption{Example of Construction with low PAR$_{IHN}$}
 \label{construction3}
\end{figure}

\begin{example}[$n=9$]
Use the triangle graph
$f = 01,02,12$ as a template.
Let $t_0 = t_1 = t_2 = 3$. (See fig.~\ref{construction3}.)
Assign the permutations
\begin{eqnarray*}
\Gamma_{0,1} = \Gamma_{0,2} &=& (12,0,1,2)(01,2)(02,1,2),\\
\Gamma_{1,0} &=& (34,5)(35,4,5)(45,3,4,5),\\
\Gamma_{1,2} &=& (45,3,4,5)(34,5)(35,4,5),\\
\Gamma_{2,0} &=& (68,7,8)(78,6,7,8)(67,8),\\
\Gamma_{2,1} &=& (78,6,7,8)(67,8)(68,7,8).
\end{eqnarray*}
Let $g_0(\boldsymbol{y_0}) = 01,02,12$, $g_1(\boldsymbol{y_1}) = 34,35,45$, 
and $g_2(\boldsymbol{y_2}) = 67,68,78$.
Then, using~(\ref{eq3}) to construct $p(\boldsymbol{x})$ we get,
$ p(\boldsymbol{x}) = 0135,$ $0178,$ $0245,$ $0267,$ $1234,$ $1268,$ $3467,$ $3568,$ $4578,$ 
$014,$ $015,$ $016,$ $017,$ $018,$ $023,$ $024,$ $025,$ $028,$ $034,$ $068,$ $125,$ 
$127,$ $128,$ $134,$ $145,$ $167,$ $168,$ $234,$ $235,$ $245,$ $267,$ $268,$ $278,$ 
$348,$ $357,$ $358,$ $378,$ $456,$ $457,$ $458,$ $468,$ $478,$ $567,$ $568,$ $578,$ 
$05,$ $07,$ $08,$ $13,$ $14,$ $17,$ $23,$ $25,$ $26,$ $28,$ $36,$ $37,$ $38,$ $46,$ 
$56,$ $58,$ $01,$ $02,$ $12,$ $34,$ $35,$ $45,$ $67,$ $68,$ $78 $.
Then $p(\boldsymbol{x})$ has PAR$_{IHN} = 10.25$.
\end{example}

The examples of our construction satisfy a low PAR$_{IHN}$.
Further work should ascertain the proper choice of permutations.
Finally, there is an even more obvious variation of construction (\ref{eq3}), suggested by
the graphs of fig.~\ref{graphs}, where the functions $g_j$ are chosen
either to be quadratic cliques or to be further ``nested'' versions of construction
(\ref{eq3}). We will report on this variation in a future paper.


\begin{thebibliography}{10}

\bibitem{Cald:Qua}
Calderbank, A.R., Rains, E.M., Shor, P.M., Sloane, N.J.A.:
\newblock Quantum error correction via codes over {GF(4)}.
\newblock IEEE Trans. Inform. Theory \textbf{44} (1998)  pp.~1369--1387.
\\\url{http://arxiv.org/quant-ph/9608006}

\bibitem{rains}
Rains, E.M., Sloane, N.J.A.:
\newblock Self-dual codes.
\newblock In Pless, V.S., Huffman, W.C., eds.: Handbook of Coding Theory.
\newblock Elsevier (1998)  177--294.
\\\url{http://arxiv.org/math/0208001}

\bibitem{Hohn:Klein}
Höhn, G.:
\newblock Self-dual codes over the {K}leinian four group.
\newblock Mathematische Annalen \textbf{327} (2003)  pp.~227--255.
\\\url{http://arxiv.org/math/0005266}

\bibitem{Hein:GrEnt}
Hein, M., Eisert, J., Briegel, H.J.:
\newblock Multi-party entanglement in graph states.
\newblock Phys. Rev. A \textbf{69} (2004).
\\\url{http://arxiv.org/quant-ph/0307130}

\bibitem{Glynn:Tome}
Glynn, D.G., Gulliver, T.A., Maks, J.G., Gupta, M.K.:
\newblock The geometry of additive quantum codes.
\newblock Submitted to Springer-Verlag (2004)

\bibitem{Slo:Seq}
Sloane, N.J.A.:
\newblock {T}he {O}n-{L}ine {E}ncyclopedia of {I}nteger {S}equences.
\newblock Web page (2004).
\\\url{http://www.research.att.com/~njas/sequences/}

\bibitem{database}
Danielsen, L.E.:
\newblock Database of self-dual quantum codes.
\newblock Web page (2004).
\\\url{http://www.ii.uib.no/~larsed/vncorbits/}

\bibitem{Sch:QG}
Schlingemann, D., Werner, R.F.:
\newblock Quantum error-correcting codes associated with graphs.
\newblock Phys. Rev. A \textbf{65} (2002).
\\\url{http://arxiv.org/quant-ph/0012111}

\bibitem{Gras:QG}
Grassl, M., Klappenecker, A., Rotteler, M.:
\newblock Graphs, quadratic forms, and quantum codes.
\newblock In: Proc. IEEE Int. Symp. Inform. Theory. (2002)  p.~45

\bibitem{Glynn:Graph}
Glynn, D.G.:
\newblock On self-dual quantum codes and graphs.
\newblock Submitted to Elect. J. Combinatorics. (2002).
\\\url{http://homepage.mac.com/dglynn/.cv/dglynn/Public/SD-G3.pdf-link.pdf}
  
\bibitem{VanD:Gr}
Van~den Nest, M., Dehaene, J., De~Moor, B.:
\newblock Graphical description of the action of local {C}lifford
  transformations on graph states.
\newblock Phys. Rev. A \textbf{69} (2004).
\\\url{http://arxiv.org/quant-ph/0308151}

\bibitem{Par:QE}
Parker, M.G., Rijmen, V.:
\newblock The quantum entanglement of binary and bipolar sequences.
\newblock In Helleseth, T., Kumar, P.V., Yang, K., eds.: Sequences and Their
  Applications, SETA'01. Discrete Mathematics and Theoretical Computer Science
  Series, Springer-Verlag (2001).
\\\url{http://arxiv.org/quant-ph/0107106}

\bibitem{Bouch:Iso}
Bouchet, A.:
\newblock Isotropic systems.
\newblock European J. Combin. \textbf{8} (1987)  pp.~231--244

\bibitem{Bouch:VNC}
Bouchet, A.:
\newblock Recognizing locally equivalent graphs.
\newblock Discrete Math. \textbf{114} (1993)  pp.~75--86

\bibitem{nauty}
McKay, B.D.:
\newblock nauty User's Guide. (2004).
\\\url{http://cs.anu.edu.au/~bdm/nauty/nug.pdf}.

\bibitem{Gull:Circ}
Gulliver, T.A., Kim, J.-L.:
\newblock Circulant based extremal additive self-dual codes over {GF(4)}.
\newblock IEEE Trans. Inform. Theory \textbf{50} (2004)  pp.~359--366

\bibitem{Gras:QECCs}
Grassl, M.:
\newblock Bounds on $d_{min}$ for additive $[[n,k,d]]$ {QECC}.
\newblock Web page (2003).
\\\url{http://iaks-www.ira.uka.de/home/grassl/QECC/TableIII.html}

\bibitem{vncbent}
Riera, C., Petrides, G., Parker, M.G.:
\newblock Generalized bent criteria for {B}oolean functions.
\newblock Technical Report 285, Dept. of Informatics, University of Bergen,
  Norway (2004).
\\\url{http://www.ii.uib.no/publikasjoner/texrap/pdf/2004-285.pdf}

\bibitem{Par:SB}
Parker, M.G.:
\newblock Generalised {S}-box nonlinearity.
\newblock NESSIE Public Document, NES/DOC/UIB/WP5/020/A. (2003).
\\\url{https://www.cosic.esat.kuleuven.ac.be/nessie/reports/phase2/SBoxLin.pdf}

\bibitem{Dan:APC}
Danielsen, L.E., Gulliver, T.A., Parker, M.G.:
\newblock Aperiodic propagation criteria for {B}oolean functions.
\newblock Submitted to Inform. Comput. (2004).
\\\url{http://www.ii.uib.no/~matthew/GenDiff4.pdf}

\bibitem{ramsey}
Radziszowski, S.P.:
\newblock Small {R}amsey numbers.
\newblock Elect. J. Combinatorics (2002)  pp.~1--42 Dynamical Survey DS1.
\\\url{http://www.combinatorics.org/Surveys/ds1.pdf}

\bibitem{interlace2}
Arratia, R., Bollob\'{a}s, B., Sorkin, G.B.:
\newblock The interlace polynomial of a graph.
\newblock J. Combin. Theory Ser. B \textbf{92} (2004)  pp.~199--233.
\\\url{http://arxiv.org/math/0209045}

\bibitem{aigner}
Aigner, M., van~der Holst, H.:
\newblock Interlace polynomials.
\newblock Linear Algebra and its Applications \textbf{377} (2004)  pp.~11--30

\bibitem{spectralinterlace}
Riera, C., Parker, M.G.:
\newblock Spectral interpretations of the interlace polynomial.
\newblock Submitted to WCC2005. (2004).
\\\url{http://www.ii.uib.no/~matthew/WCC4.pdf}

\bibitem{wheelPAR}
Parker, M.G., Gulliver, T.A.:
\newblock On graph symmetries and equivalence of the six variable double-clique
  and wheel.
\newblock Unpublished (2003)

\bibitem{Par:PAR1}
Parker, M.G., Tellambura, C.:
\newblock A construction for binary sequence sets with low peak-to-average
  power ratio.
\newblock In: Proc. IEEE Int. Symp. Inform. Theory. (2002)  p.~239.
\\\url{http://www.ii.uib.no/~matthew/634isit02.pdf}

\bibitem{Par:LowPAR}
Parker, M.G., Tellambura, C.:
\newblock A construction for binary sequence sets with low peak-to-average
  power ratio.
\newblock Technical Report 242, Dept. of Informatics, University of Bergen,
  Norway (2003).
\\\url{http://www.ii.uib.no/publikasjoner/texrap/pdf/2003-242.pdf}

\bibitem{Dav:PF}
Davis, J.A., Jedwab, J.:
\newblock Peak-to-mean power control in {OFDM}, {G}olay complementary sequences
  and {R}eed-{M}uller codes.
\newblock IEEE Trans. Inform. Theory \textbf{45} (1999)  pp.~2397--2417

\end{thebibliography}
\end{document}